\documentstyle[11pt]{article} 
\begin{document}
\hoffset = -1truecm
\voffset = -2truecm
\hyphenation{self-cons-ist-ent}
\title{STATIC EINSTEIN-MAXWELL SOLUTIONS IN 2+1 DIMENSIONS\thanks{gr-qc/9605049}}
\author{Mauricio Cataldo\thanks{e-mail: mcataldo@zeus.dci.ubiobio.cl} \\
  {\it Departamento de F\'\i sica, Facultad de Ciencias, 
           Universidad del B\'\i o-B\'\i o, Chile.}\\ 
\mbox{} \\
Patricio Salgado\thanks{e-mail: psalgado@halcon.dpi.udec.cl}\\
{\it Departamento de F\'\i sica, Facultad de Ciencias,
           Universidad de Concepci\'on, Chile.}}
\maketitle
\begin{abstract}
We obtain the Einstein-Maxwell equations for 
(2+1)-dimensional static space-time, which are invariant under the 
transformation $q_0=i\,q_2,q_2=i\,q_0,\alpha \rightleftharpoons \gamma$.
It is shown that the magnetic solution obtained with the help of the procedure
used in Ref.~\cite{Cataldo}, can be obtained from the static BTZ solution using
an appropriate transformation. Superpositions of a perfect fluid and an
electric or a magnetic field are separately studied and their corresponding
solutions found. 
\end{abstract}
It is well known that in (2+1)-dimensional space-time, the metric around a
point mass is given by
\begin{eqnarray}
\displaystyle ds^{2}=dt^{2}-\frac{dr^{2}}{(1-\frac{kM}{2\pi})^{2}}%
-r^{2}d\phi^{2}.  \label{ec.1}
\end{eqnarray}
This space-time is analogous to the Schwarzschild metric  around a point mass 
in
(3+1)-dimensions. The metric~(\ref{ec.1}) corresponds to a flat space-time~%
\cite{Gott}. Another situation in (2+1) dimensions is to consider the
cosmological constant $\Lambda$. In this case, outside sources, the exterior
gravitational fields are spaces of constant curvature: De Sitter for $%
\Lambda > 0 $ and anti-De Sitter for $\Lambda < 0 $ \cite{Deser1}.

The (2+1) local electromagnetic field is given by (``cutting out'' one of
the spatial dimensions ) 
\begin{eqnarray}  \label{T. Maxwell}
F= \frac{1}{2} \, F_{a b} \, dx^{a}\, \wedge \,dx^{b}= E_{1} \, dx^{1}\,
\wedge \,dx^{0}+ \\
E_{2} \, dx^{2}\, \wedge \,dx^{0}+ B \, dx^{1}\, \wedge \, dx^{2}. 
\nonumber
\end{eqnarray}
Thus the electromagnetic tensor has only three independent components~\cite
{Gott1,Gott}, two for the vector electric field and one for the scalar
magnetic field ($\vec{B}$ to be a vector needs the missing dimension).

On the other hand, since the tensor of Levi-Civit\'a in a three dimensional
space has the form $\varepsilon_{[a b c]}$ with $\varepsilon_{0 1 2} = 1$,
the dual of~(\ref{T. Maxwell}) is a vector given by 
$*F_{a}= \frac{1}{2} \, \varepsilon_{a b c} \, F^{b c}$,
(latin indices are assumed to take on the values 0, 1, 2). Therefore, the
source free Maxwell's equations in (2+1)-dimensional gravity lack invariance
under dual transformation. This means that, for a given Einstein-Maxwell
space-time without sources, it is not possible to transform an electric
field into a magnetic field and vice versa. So that for the self-consistent
problem we must solve the Einstein-Maxwell equations for an electric or for
a magnetic fields separately.

The presence of electromagnetic fields can curve the space outside sources.
Also, the curvature is present when the spacetime is filled with a perfect
fluid. The above picture is confirmed by the following electrovacuum static
solution, found by Gott {\em et al} \cite{Gott}: 
\begin{equation} \label{Gott}
ds^{2}=\frac{k \, Q^{2}}{2\pi} \, ln \left( \frac{r_{c}}{r} \right) \,
dt^{2}- \frac{2\pi}{k \, Q^{2}} \left( ln \left( \frac{r_{c}}{r} \right)
\right)^{-1} \, dr^{2}-r^{2}d\phi,  \label{ec.2}
\end{equation}
where $r_{c}$ and $Q$ are constants and the electric field is given by $%
E_{r}=Q/r$. This spacetime has a horizon at $r=r_{c}$ and is the analog to
the Reissner-Nordstr\"{o}m solution in (3+1)-dimensions. Recently Ba\~nados,
Teitelboim and Zanelli (BTZ) found the static electrically charged solution (%
$J=0$)~\cite{Teitelboim}, which is the 2+1 Kottler analog~\cite
{Kottler,Cataldo2} 
\begin{eqnarray}  \label{BTZ}
ds^{2}=h(r) \, dt^{2} - h^{- 1}(r) \, dr^{2} -r^{2} \, d\phi^{2},
\end{eqnarray}
where $h(r)=-M+\frac{r^{2}}{l^{2}}-\frac{1}{2} \, Q^{2} \, ln\,r $
($-\infty < t < \infty$, $0 < r < \infty$, and $0 \leq \phi \leq 2\pi$).
The constant $l$ is related to the cosmological constant $\Lambda$ by $%
\Lambda= - l^{-2}$. In order that the horizon exists, one must have $M > 0$.
If $\Lambda=0$, we obtain the solution~(\ref{ec.2}). The BTZ solution, however,
does not completely describe the coupling between the gravitational and 
electromagnetic fields in (2+1) dimensions because the magnetic case must be 
considered separately.

In this work we find the exact Einstein-Maxwell solutions which give a
complete description of the static superpositions of a perfect fluid and an
electric or a magnetic field. The solutions are obtained with the help of
the procedure used in ref.~\cite{Cataldo}, where the four-dimensional case
was studied for a neutral perfect fluid filling a spherically symmetric
space-time.

In 2+1 dimensions, the metric for an arbitrary static circularly symmetric 
space-time can be written in the form 
\begin{eqnarray}  \label{metrica}
ds^{2}= e^{2 \, \alpha(r)}\, dt^{2}- e^{2 \, \beta(r)} \, dr^{2}- e^{2 \,
\gamma(r)} \, d\theta^{2}.
\end{eqnarray}
The general form of the electromagnetic field tensor which shares the static
circularly symmetric space-time is given by 
$F= E_{r} \, dr \, \wedge \,dt + B \, dr \, \wedge \,d\theta$.
The electromagnetic field tensor in terms of a four-potential 
\begin{eqnarray}  \label{4-potencial}
A= A_{a}(r)\,dx^{a}
\end{eqnarray}
is given by $F = dA= \frac{1}{2}\,F_{a b}\,dx^{a} \wedge dx^{b}$,
where the functions $A_{a}(r)$ can be freely specified. So 
$dA=F=q_{0}\,A^{\prime}_{0}\, dr \wedge dt + q_{2}\,A^{\prime}_{2} \,dr
\wedge d\theta$
where $A_{0}$ and $A_{2}$ are arbitrary functions of the $r$ coordinate, the
differentiation with respect to $r$ is denoted by ' and the constant
coefficients $q_{0}$ and $q_{2}$ are introduced for switching off the
electric and/or magnetic fields. This implies that the $r$-component of the
electric field is given by $q_{0}\,A^{\prime}_{0}$ and of the magnetic field
by $q_{2}\,A^{\prime}_{2}$.

To write the Einstein's equations we will use the tetrad formalism and
Cartan structure equations. A convenient orthonormal basis for the metric~(%
\ref{metrica}) is 
\begin{eqnarray}  \label{tetrada}
\theta^{(0)} = e^{\alpha}\, dt , \,\,\, \theta^{(1)} = e^{\beta} \, dr, \,
\,\, \theta^{(2)} = e^{\gamma} \, d\theta.
\end{eqnarray}
To construct Einstein-Maxwell fields, we must consider the Maxwell's
equations and the stress-energy tensor of the electromagnetic field which,
with respect to~(\ref{tetrada}) in Gaussian units, is defined by ~\cite{Gott}
\begin{eqnarray}  \label{tensor electromagnetico}
T_{(a)(b)}= \frac{g_{(a)(b)}}{8 \pi} \, F_{(c)(d)} \, F^{(c)(d)} -
\frac{1}{2 \pi} F_{(a)(c)} \, F_{(b)}^{ \,\,\,\,\,\, (c)}
\end{eqnarray}
In (2+1) dimensions the trace of~(\ref{tensor electromagnetico}) is
$T=\frac{1}{8\pi} \, F_{a b} \, F^{a b}$.
To get its components we must compute $F_{(a)(b)}$. In the coordinate basis 
\begin{eqnarray}  \label{maxwellito}
F_{a b}= \frac{1}{2} \, q_{0}\,A^{\prime}_{0}\,\delta^{r}_{[a}\,
\delta^{t}_{b]}+ \frac{1}{2} \, q_{2}\,A^{\prime}_{2}\,\delta^{r}_{[a}\,
\delta^{\theta}_{b]}
\end{eqnarray}
or in the basis~(\ref{tetrada}) 
$F_{(a) (b)}= q_{0}\,I_{e} \, \delta^{1}_{[a}\, \delta^{0}_{b]} + 
q_{2} \, I_{m} \, \delta^{1}_{[a}\, \delta^{2}_{b]}$, where
$I_{e}=\frac{1}{2} \,A^{\prime}_{0}\, e^{- \alpha - \beta}$ and 
$I_{m}=\frac{1}{2} \, q_{2}\,A^{\prime}_{2} \, e^{- \beta - \gamma}$.
Thus from~(\ref{tensor electromagnetico}) we get 
\begin{eqnarray}
T_{(a)(b)}^{e.m.} \, \theta^{(a)} \, \otimes \, \theta^{(b)}= \left( 
\frac{q_{0}{}^{2}}{ \pi} \, I_{e}{}^{2} + \frac{q_{2}{}^{2}}{ \pi} \, 
I_{m}{}^{2} \right) \, \theta^{(0)} \, \otimes \, \theta^{(0)}-  \nonumber \\
\left( - \frac{q_{0}{}^{2}}{ \pi} \, I_{e}{}^{2} + \frac{q_{2}{}^{2}}{ \pi} \, 
I_{m}{}^{2} \right) \, \theta^{(1)} \, \otimes \, \theta^{(1)}- 
\nonumber \\
\left( \frac{q_{0}{}^{2}}{ \pi} \, I_{e}{}^{2} + \frac{q_{2}{}^{2}}{ \pi} \, 
I_{m}{}^{2} \right) \, \theta^{(2)} \, \otimes \, \theta^{(2)}- 
\nonumber \\
\frac{2 q_{0} \, q_{2}}{ \pi} I_{e} \, I_{m} \left( \theta^{(0)} \, \otimes \, 
\theta^{(2)} + \theta^{(2)} \, \otimes \, \theta^{(0)} \right). \hspace{1cm} 
\nonumber
\end{eqnarray}
From the tetrad~(\ref{tetrada}) and using Cartan exterior forms calculus the
following non-trivial components of the Einstein's equations with the 
cosmological constant are obtained:
\begin{eqnarray}  \label{cero-cero}
e^{- 2 \beta} \left( \gamma {}^{\prime} \beta {}^{\prime} -
\gamma {}^{\prime \prime}- \gamma {}^{\prime}{}^{2}\right) = \Lambda + 
\frac{\kappa q_{0}{}^{2} I_{e}{}^{2}}{\pi} + 
\frac{\kappa q_{2}{}^{2} I_{m}{}^{2}}{\pi} 
\end{eqnarray}
\begin{eqnarray}  \label{uno-uno}
- \alpha\, ^{\prime}\, \gamma\, ^{\prime}\,e^{- 2\, \beta}= \Lambda + 
\frac{\kappa q_{0}{}^{2} I_{e}{}^{2}}{\pi} - 
\frac{\kappa q_{2}{}^{2} I_{m}{}^{2}}{\pi} 
\end{eqnarray}
\begin{eqnarray}  \label{dos-dos}
e^{- 2 \beta} \left( \alpha {}^{\prime} \beta {}^{\prime}- 
\alpha {}^{\prime\prime}- \alpha {}^{\prime}{}^{2} \right) = \Lambda - 
\frac{\kappa q_{0}{}^{2} I_{e}{}^{2}}{\pi} -
\frac{\kappa q_{2}{}^{2} I_{m}{}^{2}}{\pi} 
\end{eqnarray}
\begin{eqnarray}  \label{cero-dos}
\frac{2 \kappa \, q_{0} \, q_{2}}{\pi} \, I_{e} \, I_{m} = 0
\end{eqnarray}

Now we must consider the Maxwell's equations. The contravariant density
components of~(\ref{maxwellito}) are 
\begin{eqnarray*}
\sqrt{-g} \, F^{a b}=e^{\alpha+\beta+\gamma} \left(\frac{2 q_{0} I_{e}{}^{2}}
{A^{\prime}_{0}} \, \delta^{[a}_{r} \, \delta^{b]}_{t} + 
\frac{2 q_{2} I_{m}{}^{2}} {A^{\prime}_{2}} \, \delta^{[a}_{r} \, 
\delta^{b]}_{\theta} \right)
\end{eqnarray*}
It is clear that the source-free Maxwell's equations are satisfied if
\begin{eqnarray}  \label{ecuacion electro-magnetica}
e^{ \alpha + \beta - \gamma}= A^{\prime}_{0}, \,\,
e^{- \alpha + \beta + \gamma}= A^{\prime}_{2}
\end{eqnarray}
where the constants of integration, without any loss of generality, have
been made equal to~$1$. So the Einstein-Maxwell equations are given by Eqs.~(
\ref{cero-cero})-(\ref{cero-dos}) and~(\ref{ecuacion electro-magnetica}). To 
obtain the Einstein-Maxwell solutions it is useful
to notice that Eq.~(\ref{cero-dos}) says to us that either the electric or
the magnetic field must be zero.

It is easy to check that under the transformation 
\begin{eqnarray}  \label{transformacion1}
q_{0}= i \, q_{2}, \,\,\,\,\,\,\, q_{2}= i \, q_{0}, \,\,\,\,\,\,\, \alpha
\rightleftharpoons \gamma
\end{eqnarray}
the Einstein-Maxwell equations are invariant. This \\
means that if we have the
magnetic solution, then one can obtain the electric analog by making the
formal transformation~(\ref{transformacion1}). In other words, if the
Einstein-Maxwell solution is given in the form~(\ref{metrica}), then we
obtain an analogous metric making 
\begin{eqnarray}  \label{transformacion2}
dt= i \, d\theta, \,\,\,\,\,\,\,\,\,\, d\theta= i \, dt,
\,\,\,\,\,\,\,\,\,\, q_{0}= i \, q_{2}, \,\,\,\,\,\,\,\,\,\, q_{2}= i \,
q_{0}.
\end{eqnarray}


Now, we will find a general magnetic solution (in analytical form). So
we must consider Eq.~(\ref{cero-dos}). That means that in this case either
the electric or the magnetic field must be zero, so that in the considered
equations we must set $q_{0}= 0$ (or $A_{0}= 0$). This implies that, in
order to solve the self-consistent equations, it is not necessarily to
consider the first condition of~(\ref{ecuacion electro-magnetica}). 
Subtracting equations (\ref{uno-uno}) and (\ref{dos-dos}) and using the second
condition of~(\ref{ecuacion electro-magnetica}), we obtain 
$e^{2 \alpha} = D\, e^{C\, A_{2}(r)}$,
where $A_{2}$ is an arbitrary function of $r$, and $C$ and $D$ are constants
of integration. On the other hand the combination of~(\ref{uno-uno}) and~(%
\ref{dos-dos}) leads us to the equation 
\begin{eqnarray}
\left(\alpha\,^{^{\prime}}\, e^{\alpha - \beta+ \gamma} \right)^{^{\prime}}=
- 2 \, \Lambda \, A^{^{\prime}}_{2} \, D \, e^{C \, A} + \frac{\kappa}{2 \pi}%
\, q_{2}{}^{2}\, A^{\prime}_{2}
\end{eqnarray}
which yields with the help of~(\ref{ecuacion electro-magnetica}) the following
expression for the function $e^{2 \gamma}$: 
\begin{eqnarray}
e^{2 \gamma}= \frac{\kappa}{\pi \, C}\, q_{2}{}^{2}\, A_{2}- \frac{4 \,
\Lambda \, D}{C^{2}}\, e^{C \, A_{2}} + F,
\end{eqnarray}
where $F$ is a new constant of integration. From~(\ref{ecuacion electro-magnetica})
we obtain 
$e^{2 \beta}= D\, A^{\prime}_{2} \, ^{2}\, e^{C \, A_{2}} \, e^{-2 \gamma}$.
Introducing a new coordinate $\tilde{r}$ defined by $\tilde{r}= A_{2}(r)$,
we have
\begin{eqnarray}  \label{metrica2}
ds^{2} = D e^{C\, \tilde{r}} dt^{2}- D e^{C\, \tilde{r}} \,e^{-2 \gamma}
 d\tilde{r}^{2} - e^{2 \gamma} d\theta^{2}. 
\end{eqnarray}
where now $e^{2 \gamma}=\frac{\kappa \, q_{2}{}^{2}}{\pi \, C}\, \tilde{r}- 
\frac{4 \, \Lambda \, D}{C^{2}}\, e^{C \, \tilde{r}} + F$.
In this coordinate gauge the magnetic field is constant. The 
solution~(\ref{metrica2}) can not be carried out in the spatial gauge 
$g_{2 2}= r^{2}$ with the help of the transformation 
\begin{eqnarray}  \label{gauge}
\tilde r^{2} = \frac{\kappa \, q_{2}{}^{2}}{\pi \, C}\, r -\frac{4 \,
\Lambda \, D}{C^{2}}\, e^{C \, r} + F,
\end{eqnarray}
because the resulting metric does not take an analytical form. However it is
possible to obtain the metrics for the gauge $g_{2 2}= r^{2}$ in exact form
by setting $\Lambda=0$ or switching off the magnetic field ($q_{2}=0$). This
means that if $q_{2}=0$ then, from~(\ref{gauge}) one can obtain the BTZ
non-charged three-dimensional black hole~\cite{Teitelboim}; and that when $%
\Lambda=0$ we obtain the 3+1 magnetic Reissner-Nordstr\"om counterpart. In
fact, in this case we obtain the following metric: 
\begin{eqnarray}  \label{Reissner2+1}
ds^{2}= e^{a \, r^{2}} dt^{2}- e^{a \, r^{2}} dr^{2}- r^{2} d\theta^{2},
\end{eqnarray}
with $a=\frac{\kappa \, q_{2}{}^{2}}{\pi}$. From Eq.~(\ref{Reissner2+1}) it
follows that the 2+1 magnetic monopole counterpart is not a black hole,
contrasting with the 2+1 electric analog. If we switch off the magnetic
field ($a=0$), one the flat three-dimensional space-time gets.

Finally, we note that application of transformation (\ref{transformacion2})
on~(\ref{metrica2}) leads to 
\begin{eqnarray}  \label{metricaBTZ}
ds^{2}= e^{2 \alpha} dt^{2} - D e^{C \, \tilde{r}} e^{-2 \alpha} \, 
d\tilde{r}^{2}- D \, e^{C\, \tilde{r}} \, d\theta^{2}.
\end{eqnarray}
where $e^{2 \alpha}=\frac{- \kappa \, q_{2}{}^{2}}{\pi \, C}\, \tilde{r} - 
\frac{4 \, \Lambda \, D}{C^{2}}\, e^{C \, \tilde{r}} + F$. In the gauge 
$g_{2 2}=r^{2}$ the metric~(\ref{metricaBTZ}) takes on the new form 
\begin{eqnarray}  \label{metricaBTZBTZ}
ds^{2}= e^{2 \alpha} dt^{2} - e^{-2 \alpha} d{r}^{2} - r^{2} \, d\theta^{2},
\end{eqnarray}
where now $e^{2 \alpha}=\frac{- 2 \kappa \, q_{2}{}^{2}}{\pi} \, 
ln \, r - \Lambda \, r^{2} + F$. Then the BTZ charged black hole~(\ref{BTZ}) is
obtained if $\Lambda= - l^{-2}$, $F=-M$ and 
$Q^{2}= 4 \kappa \, q_{2}{}^{2}/ \pi$. When $\Lambda=0$ one gets the 
electrically charged solution~(\ref{Gott}). In this case the
electromagnetic potential is given by $A= ln \, r$. This means that 
$dA=q_{2}/r \, dr \wedge dt$ (or $E_{r}=q_{2} /r$). If one uses the
transformation~(\ref{transformacion2}), then $dA= - q_{0}/r \, dr \wedge
d\theta$, so that the electric field is replaced by a magnetic field and $%
B=- q_{0}/r$.

Now, in order to obtain the Einstein-Maxwell fields with a neutral fluid, we 
must consider the stress-energy tensor of the perfect fluid, which is given by 
\begin{eqnarray}  \label{fluido}
T_{(a)(b)}^{p.f.} = (\mu +p) \, U_{(a)} U_{(b)} - p \, g_{(a)(b)},
\end{eqnarray}
where $\mu$ and $p$ are the mass-energy density and pressure of the fluid,
respectively. $U_{(a)}$ is its timelike 4-velocity. If we take the four
velocity $ {\bf U} = \theta^{(0)}, $ then~(\ref{fluido}) becomes 
$T_{(0)(0)}^{p.f.}=\mu$, $T_{(1)(1)}^{p.f.}=T_{(2)(2)}^{p.f.}=
T_{(3)(3)}^{p.f.}=p$. With the electromagnetic potential~(\ref{4-potencial}) 
and a neutral perfect fluid, the Einstein-Maxwell equations with cosmological
constant are now given by 
\begin{eqnarray}  \label{rn-cero-cero}
e^{-2\beta} \left(\gamma {}^{\prime} \beta {}^{\prime} -
\gamma {}^{\prime \prime}- \gamma {}^{\prime}{}^{2}\right) = \Lambda + 
\frac{\kappa q_{0}{}^{2} I_{e}{}^{2}}{\pi} + \\
\frac{\kappa q_{2}{}^{2} I_{m}{}^{2}}{\pi} + \kappa \mu \hspace{2cm}
\nonumber
\end{eqnarray}
\begin{eqnarray}  \label{rn-uno-uno}
- \alpha\, ^{\prime}\, \gamma\, ^{\prime}\,e^{- 2\, \beta}= \Lambda + 
\frac{\kappa q_{0}{}^{2} I_{e}{}^{2}}{\pi} - 
\frac{\kappa q_{2}{}^{2} I_{m}{}^{2}}{\pi} - \kappa p
\end{eqnarray}
\begin{eqnarray}  \label{rn-dos-dos}
e^{- 2 \beta} \left( \alpha {}^{\prime} \beta {}^{\prime}- 
\alpha {}^{\prime\prime}- \alpha {}^{\prime}{}^{2} \right) = \Lambda - 
\frac{\kappa q_{0}{}^{2} I_{e}{}^{2}}{\pi} - \\
\frac{\kappa q_{2}{}^{2} I_{m}{}^{2}}{\pi} - \kappa p \hspace{2cm}
\nonumber
\end{eqnarray} 
\begin{eqnarray}  \label{rn-cero-dos}
\frac{2 \kappa \, q_{0} \, q_{2}}{\pi} \, I_{e} \, I_{m} = 0
\end{eqnarray}
and by conditions~(\ref{ecuacion electro-magnetica}). It is easy to see that 
equations~(\ref{rn-cero-cero})-(\ref{rn-cero-dos})
are not invariant under transformation (\ref{transformacion2}). This means
that one must solve the system~(\ref{rn-cero-cero})-(\ref{rn-cero-dos}) for
an electric or a magnetic field separately (Eq.~(\ref{rn-cero-dos}) implies
that either the electric or the magnetic field must be zero).

First we consider the electric solution: In this case we can not obtain the
solution for the general electromagnetic potential~(\ref{4-potencial}) with
an arbitrary function $A_{0}$ ($A_{2}=0$). We must consider some concrete
function, such as the 4-potential in the following form 
\begin{eqnarray}
A_{0} =\left\{ 
\begin{array}{ll}
\frac{1}{n+1} \, r^{n+1} \, dt \,\,\, & \mbox {if $n\neq-1$} \\ 
ln\, r \, dt \,\,\, & \mbox {if $n=-1$}
\end{array}
\right.  \label{rn-potencial}
\end{eqnarray}
where $n$ is an arbitrary constant, and solve the Einstein-maxwell 
equations.
Thus the electric field takes the form $E=q_{0} \, r^{n}$. In the case of
the circularly symmetric metric, one takes $e^{2 \, \gamma}= r^{2}$,
arriving at the following metrics:~\cite{Firenze} \\
For {\bf $n \neq 1, -1, -3$}, 
\begin{eqnarray*}
\lefteqn{e^{2\alpha}= r^{2(n+1)} e^{- 2 \beta}=
 \frac{\kappa q^{2} r^{2(n+1)}}{4 \pi (n-1)(n+1)}  +\frac{A r^{n+3}}{n+3} + B} \\
 & & \kappa p=\frac{\kappa q^{2}(n+1)}{8\pi (n-1)r^{2}} + \frac{A}{2}r^{-(n+1)} +
\Lambda, \\
 & & \kappa \mu=\frac{B(n+1)}{r^{2(n+2)}}+\frac{A(n-1)}{2(n+3)}r^{-(n+1)}- \frac{\kappa
q^{2}}{8 \pi r^{2}}- \Lambda.
\end{eqnarray*}
For {\bf $n=-1$},
\begin{eqnarray*}  
\lefteqn{e^{2\alpha}=e^{-2 \beta}=-\frac{\kappa q^{2}}{4\pi}ln\,r + Ar^{2}+B,}
 \\
 & & \kappa p = -\kappa \mu = A + \Lambda.
\end{eqnarray*}
For {\bf $n=1$}, 
\begin{eqnarray*}
\lefteqn{e^{2\alpha}= r^{4} e^{-2 \beta}= \frac{\kappa q^{2}}{8\pi}r^{4}\,ln\,r + A r^{4} + B,}
\\
 & & \kappa p= \frac{\kappa q^{2}}{16\pi r^{2}}(3+4ln\,r)+ \frac{2A}{r^{2}} +
\Lambda, \\
 & & \kappa \mu= -\frac{3\kappa q^{2}}{16\pi r^{2}} + \frac{2B}{r^{6}} - 
\Lambda.
\end{eqnarray*}
For {\bf $n=-3$}, 
\begin{eqnarray*}
\lefteqn{e^{2\alpha}= r^{-4} e^{-2 \beta}=
\frac{\kappa q^{2}}{32\pi r^{4}} + A ln\,r + B,}\\
 & & \kappa p= \frac{\kappa q^{2}}{16\pi r^{2}} + \frac{A}{2}r^{2} + \Lambda,\\
 & & \kappa \mu= -\frac{\kappa q^{2}}{8\pi r^{2}}- \frac{Ar^{2}}{2}(1+ln\,r)-
2Br^{2} - \Lambda.
\end{eqnarray*}
From $n=-1$ we see that if $A = - \Lambda$ then we obtain the 2+1
Kottler solution analog~(\ref{BTZ}). We remark that when $B=q=0$ and the
fluid obeys a $\gamma$-law equation, i.e., $\mu$ and $p$ are related by an
equation of the form $p=(\gamma-1)\mu$ where $\gamma$ is a constant (which,
for physical reasons satisfies the inequality $1 \leq \gamma \leq 2$),  the
constant $\gamma$ may be expressed as $  \gamma=2 \, \frac{n+1}{n-1}, $
where the limits of $n$ are $-\infty \leq n \leq -3$. In this case 
\begin{equation}
\kappa \mu =\kappa \frac{n-1}{n+3}p=\frac{A(n-1)}{2(n+3)}r^{-(n+1)}.
\end{equation}
Recently, G\"urses~\cite{Gurses} obtained a class of metrics of Einstein
theory with perfect fluid sources in 2+1 dimensions. However this class of 
solutions was found for the particular case $\mu=$const and $p=$const.

Finally, we present the circularly symmetric magnetic case ($A_{0}=0$) which
takes the following form: 
\begin{eqnarray*}
ds^{2}=D\, e^{C \, A_{2} (r)} \left ( dt^{2} - r^{-2} \, A^{\prime}_{2}{}^{2} 
\, dr^{2} \right) - r^{2} \, d \theta^{2},
\end{eqnarray*}
where the mass-energy density is given by 
\begin{eqnarray*}
\kappa \, \mu = \frac{e^{- C \, A_{2}}}{ D}\, \left( \frac{C \, r}{2 \,
A^{\prime}_{2}} + \frac{A^{\prime\prime}_{2}}{A^{\prime}_{2}{}^{3}} - \frac{1%
}{A^{\prime}_{2}{}^{2}} - \frac{\kappa \, q_{2}^{2}}{4 \, \pi} \right) -
\Lambda,
\end{eqnarray*}
and the pressure by 
\begin{eqnarray*}
\kappa \, p = \frac{e^{- C \, A_{2}}}{ D} \, \left( \frac{C \, r}{2 \,
A^{\prime}_{2}} - \frac{\kappa \, q_{2}^{2}}{4 \, \pi} \right) + \Lambda.
\end{eqnarray*}

Before we finish this paper we would like to make a few comments about our
most important results. We have studied the static Einstein-Maxwell fields in
(2+1)-dimensions and obtained new exact solutions with circular symmetry.
All of them were found in the presence of the cosmological constant $\Lambda$%
. It is noteworthy that the 2+1 magnetic Reissner-Nordstr\"om analog is not
a black hole in contrast with the 2+1 electric Reissner-Nordstr\"om analog,
where a black hole is present. It is shown that the magnetic solution
obtained with the help of the procedure used in~\cite{Cataldo}, can be
obtained from the static electrically charged BTZ metric using
transformation~(\ref{transformacion2}). In this case the radial parameter
satisfies $0 < r < \infty$.

Superpositions of a perfect fluid and an electric or a magnetic field are
separately studied and their corresponding solutions found.

Shortly after we completed this work M. Ba\~nados informed us of the
existence of ref.~\cite{Welch} where a magnetic solution (HW solution) was
also obtained by a different procedure. However, this solution uses a gauge
which lacks a parameter. This could be seen by introducing the new coordinate
(and taking a negative cosmological constant $\Lambda= - l^{- 2}$) 
$\tilde{r}= ln \left( \frac{r^{2}/ l^{2}- M}{D} \right)^{1/C}$.

Then~(\ref{metrica2}) with this new coordinate becomes \\
$ds^{2}= \left( \frac{r^{2}}{l^{2}}- M \right) dt^{2} - r^{2} 
\left(\frac{r^{2}}{l^{2}}- M \right)^{-1} e^{- 2 \gamma} dr^{2} - 
e^{2 \gamma}d\theta^{2}$, \\
where $e^{2 \gamma}=r^{2} + \frac{\kappa \, q_{2}{}^{2} \, l^{4}}{4 \, \pi} 
\, ln \left(r^{2}/ l^{2}- M \right) + F $. In this gauge then the lacking 
parameter is $F$.




We would like to thank P. Minning for carefully reading the manuscript. This
work was supported in part by Direcci\'on de Promoci\'on y Desarrollo de la
Universidad del B\'\i o-B\'\i o through Grants $No.$ 951105-1 and 
$No.$ 960205-1, and in part by Direcci\'on de Investigaci\'on de la Universidad
de Concepci\'on through Grant $\# 94.11.09-1$. 


\end{document}